\documentclass[%
reprint,
superscriptaddress,
bibnotes,
 amsmath,amssymb,
prb,
longbibliography
]{revtex4-1}
\usepackage{float}
\usepackage[caption = false]{subfig}
\usepackage[final]{graphicx}

\usepackage{amssymb,amsmath}
\usepackage{color}
\usepackage{longtable}
\usepackage{soul}
\usepackage{graphicx}% Include figure files
\usepackage{epstopdf}
\usepackage{dcolumn}% Align table columns on decimal point
\usepackage{bm}% bold math
\usepackage{hyperref}% add hypertext capabilities
\usepackage{tabularx}
\usepackage[dvipsnames]{xcolor}
\usepackage{soul}
\usepackage{lineno, blindtext, color}
\begin{document}

\preprint{APS/123-QED}

\title{Continuous magnetic phase transition in artificial square ice}% Force line breaks with \\

\author{Oles Sendetskyi}
 \affiliation{%
Laboratory for Mesoscopic Systems, Department of Materials, ETH Zurich, 8093 Zurich, Switzerland
}%
\affiliation{%
Laboratory for Multiscale Materials Experiments, Paul Scherrer Institute, 5232 Villigen PSI, Switzerland
}%

\author{Valerio Scagnoli}
\email{valerio.scagnoli@psi.ch}
\affiliation{%
Laboratory for Mesoscopic Systems, Department of Materials, ETH Zurich, 8093 Zurich, Switzerland
}%
\affiliation{%
Laboratory for Multiscale Materials Experiments, Paul Scherrer Institute, 5232 Villigen PSI, Switzerland
}%

\author{Na$\mathrm{\ddot{e}}$mi Leo}
\affiliation{%
Laboratory for Mesoscopic Systems, Department of Materials, ETH Zurich, 8093 Zurich, Switzerland
}%
\affiliation{%
Laboratory for Multiscale Materials Experiments, Paul Scherrer Institute, 5232 Villigen PSI, Switzerland
}%

\author{Luca Anghinolfi}
\affiliation{%
Laboratory for Mesoscopic Systems, Department of Materials, ETH Zurich, 8093 Zurich, Switzerland
}%
\affiliation{%
Laboratory for Multiscale Materials Experiments, Paul Scherrer Institute, 5232 Villigen PSI, Switzerland
}%
\affiliation{%
Laboratory for Neutron Scattering and Imaging, Paul Scherrer Institute, 5232 Villigen PSI, Switzerland
}%

\author{Aurora Alberca}
\affiliation{%
Swiss Light Source, Paul Scherrer Institute, 5232 Villigen PSI, Switzerland
}%
\affiliation{%
University of Fribourg, Department of Physics and Fribourg Centre for Nanomaterials, Chemin du Musee 3, 1700 Fribourg, Switzerland
}%

\author{Jan L$\mathrm{\ddot{u}}$ning}
\affiliation{%
Sorbonne Universit\'{e}s, UPMC Univ Paris 06, UMR 7614, LCPMR, 75005 Paris, France
}%
\affiliation{%
CNRS, UMR 7614, LCPMR, 75005 Paris, France
}%

\author{Urs Staub}
\affiliation{%
Swiss Light Source, Paul Scherrer Institute, 5232 Villigen PSI, Switzerland
}%

\author{Peter Michael Derlet}
\email{peter.derlet@psi.ch}
\affiliation{%
Condensed Matter Theory Group, Paul Scherrer Institute, 5232 Villigen PSI, Switzerland
}%

\author{Laura Jane Heyderman}
\affiliation{%
Laboratory for Mesoscopic Systems, Department of Materials, ETH Zurich, 8093 Zurich, Switzerland
}%
\affiliation{%
Laboratory for Multiscale Materials Experiments, Paul Scherrer Institute, 5232 Villigen PSI, Switzerland
}%

\date{\today}% It is always \today, today,

\begin{abstract}
Critical behavior is very common in many fields of science and a wide variety of many-body systems exhibit emergent critical phenomena. The beauty of critical phase transitions lies in their scale-free properties, such that the temperature dependence of physical parameters of systems differing at the microscopic scale can be described by the same generic power laws. In this work we establish the critical properties of the antiferromagnetic phase transition in artificial square ice, showing that it belongs to the two-dimensional Ising universality class, which extends the applicability of such concepts from atomistic to mesoscopic magnets. Combining soft x-ray resonant magnetic scattering experiments and Monte Carlo simulations, we characterize the transition to the low temperature long range order expected for the artificial square ice system. By measuring the critical scattering, we provide direct quantitative evidence of a continuous magnetic phase transition, obtaining critical exponents which are compatible with those of the two-dimensional Ising universality class. In addition, by varying the blocking temperature relative to the phase transition temperature, we demonstrate its influence on the out-of-equilibrium dynamics due to critical slowing down at the phase transition.
\end{abstract}

\pacs{Valid PACS appear here}% PACS, the Physics and Astronomy
\maketitle

%\tableofcontents

\section{\label{sec:introduction}Introduction}

Frustrated magnetic systems have been a fertile playground to study spin ice\cite{tabata2006,fennell2007} and spin liquid\cite{canals1998,balents2010} physics, as well as emergent magnetic coulomb phases.\cite{brooksbartlett2014,henley2010,fennell2009} Recently, artificial spin systems, consisting of elongated single-domain ferromagnetic nanomagnets\cite{wang2006,nisoli2013,heyderman2013} placed on the nodes of two-dimensional lattices and coupled via their dipolar magnetic fields, have also been used to address open questions in frustrated magnetism. These systems provide the freedom to tailor the lattice geometry, as well as to observe magnetic moment configurations with microscopy techniques,\cite{farhan2013, farhanprl2013,kapaklis2014,morgan2011,perrin2016, morgan2013,qi2008,wang2006} offering great flexibility for creating and testing of novel two-dimensional magnetic systems. The most notable examples are artificial square ice and kagome ice that were initially designed as a two dimensional analog to water and pyrochlore spin ice, and are named after the lattice they are based on.\cite{wang2006,moller2006} 

Due to the direct and novel physical insight they provide, artificial spin ices have stimulated intense theoretical and experimental work. For example, it has been demonstrated numerically that the dipolar interacting kagome and square artificial spin systems admit low temperature ordered phases.\cite{moller2006,moller2009,chern2011,chern2012} One of the major developments in the field that provided the possibility to test these predictions was the creation of artificial spin ice with thermally active (superparamagnetic) nanomagnets.\cite{kapaklis2012,arnalds2012apl,morley2017} Indeed, by carefully varying the volume of the nanomagnets one can change the onset of superparamagnetic behaviour. As a result, an experimental indication of collective phase transitions to ordered phases in artificial kagome spin ice has been reported {\color{black}by Anghinolfi \it{et al.}},\cite{anghinolfi2015} and one of the open questions left to address is whether or not these artificial spin ice systems carry quantitative signatures of the critical behavior associated with the predicted continuous phase transition to low temperature long range order. 

Near such a continuous phase transition, properties of systems differing at the microscopic scale can be described by unifying models, which has led to the development of the concept known as universal behaviour of phase transitions.\cite{kadanoff1967, griffiths1970} When two different systems, for example a gas and a magnet, have transitions with the same properties, they are said to be in the same universality class. Scattering techniques have played an important role in testing this concept, in particular, providing the possibility to measure the critical scattering that originates from large fluctuations of the order parameter occurring in the vicinity of a phase transition, which is important for the determination of the universality class. {\color{black}In this respect, artificial square ice is a perfect system for testing critical properties of artificial magnets, as it features a well-established antiferromagnetic low temperature ordered state.\cite{morgan2011,zhang2013,porro2013, farhan2013,perron2013} However, besides a few theoretical studies,\cite{silva2012,levis2013} there have been no quantitative experimental observations of the critical properties of artificial square ice and, from the theoretical works, it is still not obvious whether there should be a continuous phase transition in such a system at all.\cite{silva2012} In this work, we examine the critical properties of artificial square ice in the vicinity of the antiferromagnetic phase transition by means of soft x-ray resonant magnetic scattering and find the clear emergence of magnetic criticality in artificial square ice. We demonstrate that complex many-body phenomena, usually associated with atomistic systems, are also inherent to nanomagnetic systems and can be investigated with modern synchrotron scattering methods.}

The paper is organized as follows: in Section~\ref{sec:sample} and \ref{sec:rxd} we describe the sample preparation and the soft x-ray resonant magnetic scattering technique used to measure critical properties across the phase transition, respectively. 
In Section~\ref{sec:nature} we present the experimental results and the Monte Carlo simulations that provide quantitative evidence of a continuous phase transition in artificial square ice. Both experimental results and finite size scaling collapse from Monte Carlo simulations confirm that the phase transition belongs to the two-dimensional Ising universality class. Finally, in Section~\ref{sec:effects} we exploit the possibility to tune the blocking temperature of the nanomagnets, a unique feature of artificial spin systems. We show how the value of the blocking temperature affects the phase transition {\color{black}and we provide guidelines for the choice of sample parameters to observe the critical properties at phase transitions.}
%%%%%%%%%%%%%%%%%%%%%%%%%%

\section{\label{sec:sample}Sample preparation}

\begin{figure}[t]
	\includegraphics[width=0.8\linewidth]{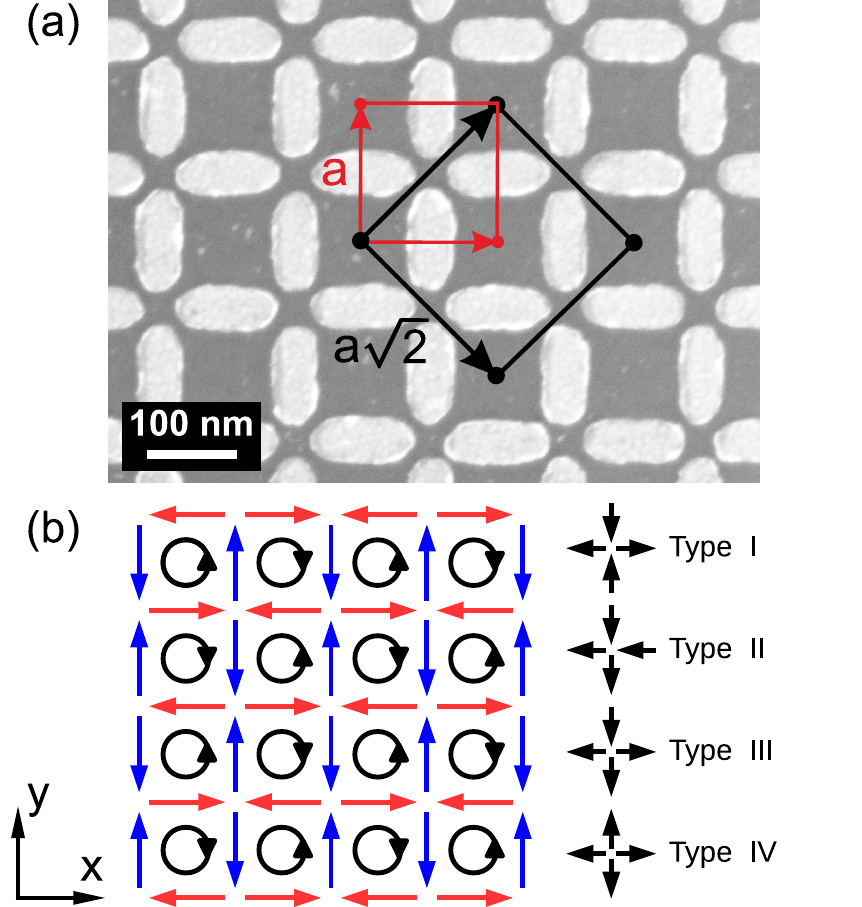}
	\caption{(a) Scanning electron microscopy image of a representative part of the artificial square ice array. The red arrows represent the structural lattice vectors of length $a$ and the black arrows indicate the magnetic lattice vectors of the ground state ordering of length $a\sqrt{2}$. (b) Schematic of the antiferromagnetic ground state reflecting the long range order of the low-temperature phase (left panel), which consists of two two-dimensional Ising systems with orthogonal moments (indicated by the red and blue arrows). This in turn may be viewed as a tile of alternating micro-vortices. The right panel lists the vertex-type convention, permutations of which produce the 16 possible vertex configurations.}
	\label{fig:figure1}
\end{figure}

The sample consists of elongated ferromagnetic nanomagnets coupled via their dipolar magnetic fields and placed on the nodes of the square lattice with a periodicity of $a=168$~nm [see Fig.~\ref{fig:figure1}(a)]. 
The nanomagnets have an elongated shape with a length $l=118$~nm and a width $w=47$~nm. Several 150$\times$1500~$\mu m^2$ arrays of nanomagnets with different thicknesses were produced on the same substrate using electron beam lithography. For this, a 70~nm-thick polymethylmethacrylate (PMMA) layer was spin-coated on a Si~(100) substrate. The patterns were exposed in the resist using a Vistec EBPG electron beam writer operated at 100~keV accelerating voltage. After development, a Permalloy (Ni$_{80}$Fe$_{20}$) film whose thickness varied across the sample was deposited by thermal evaporation as in Ref.~\onlinecite{farhan2013}. The resulting arrays had thicknesses $h$ ranging from 1 to 5~nm over a distance of 10 mm, giving a thickness change of about 0.2~nm between the neighboring arrays. The Permalloy film was then capped with a 2.5~nm-thick film of Al to prevent oxidation. The remaining resist with unwanted metallic material was removed by ultrasound-assisted lift-off in acetone. The variable thickness of the sample permitted selection of three arrays that had their magnetic transition temperatures in the experimentally accessible range. The Permalloy film thicknesses of the three arrays were measured using atomic force microscopy to be 4.1~$\pm$0.1~nm (Array~1), 4.3~$\pm$0.1~nm (Array~2) and  4.5~$\pm$0.1~nm (Array~3).

%==== figure =============================%
%=== end figure ==========================%

%==== figure =============================%
      \begin{figure}[b]
\includegraphics[width=0.99\linewidth]{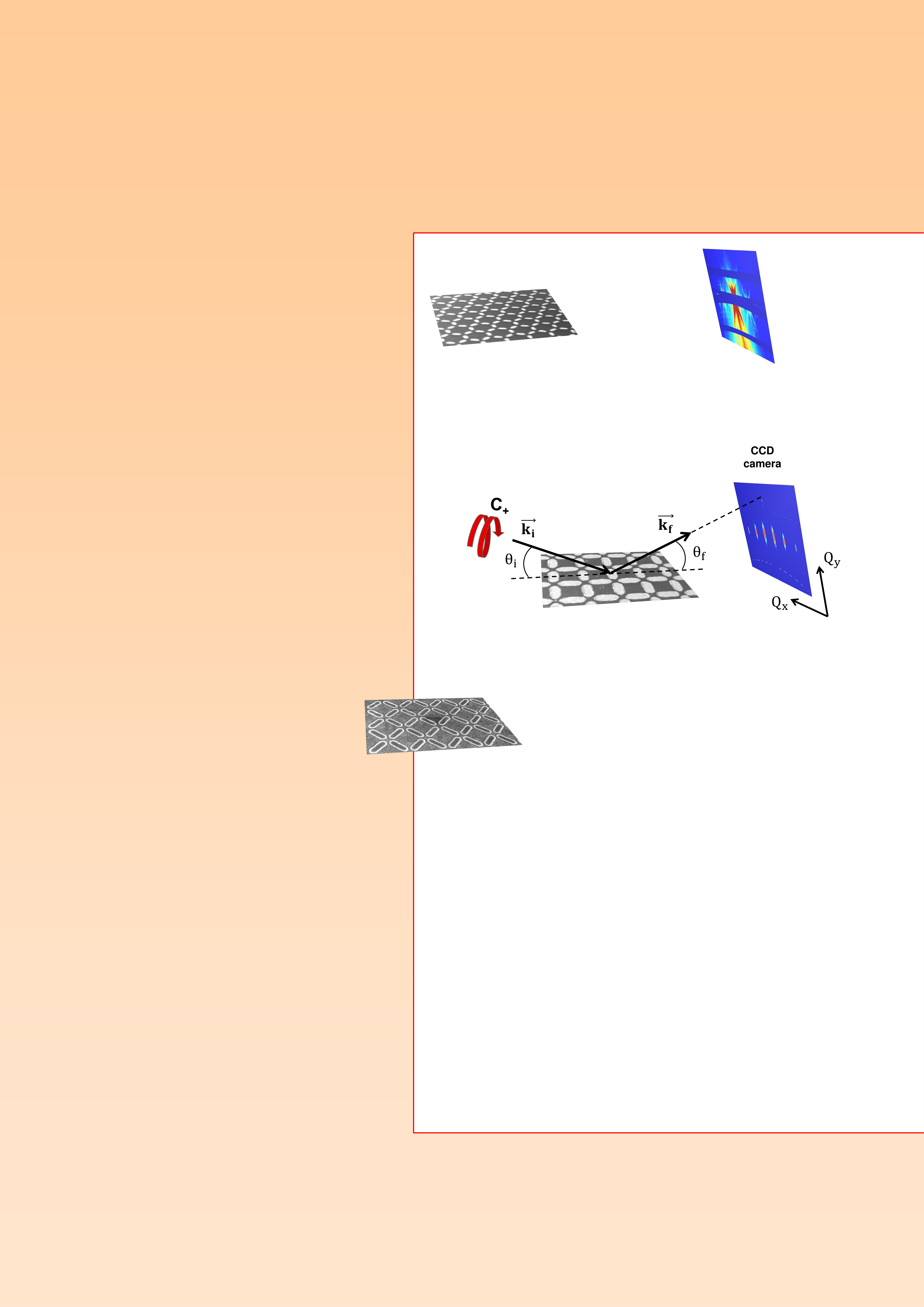} 
       \caption{Schematic illustration of the experimental geometry for soft x-ray resonant magnetic scattering experiments with circularly polarized $C_+$ synchrotron x-rays. $k_i$ ($k_f$) and $\theta_i$ ($\theta_f$) are the incident (final) x-ray wave vector and angle. $Q_x$ and $Q_y$ are the momentum transfer along the x and y directions, respectively.}
       \label{fig:geometry}
      \end{figure}
%=== end figure ==========================%

%==== figure =============================%
\begin{figure}[t]
	\includegraphics[clip,width=0.99\linewidth]{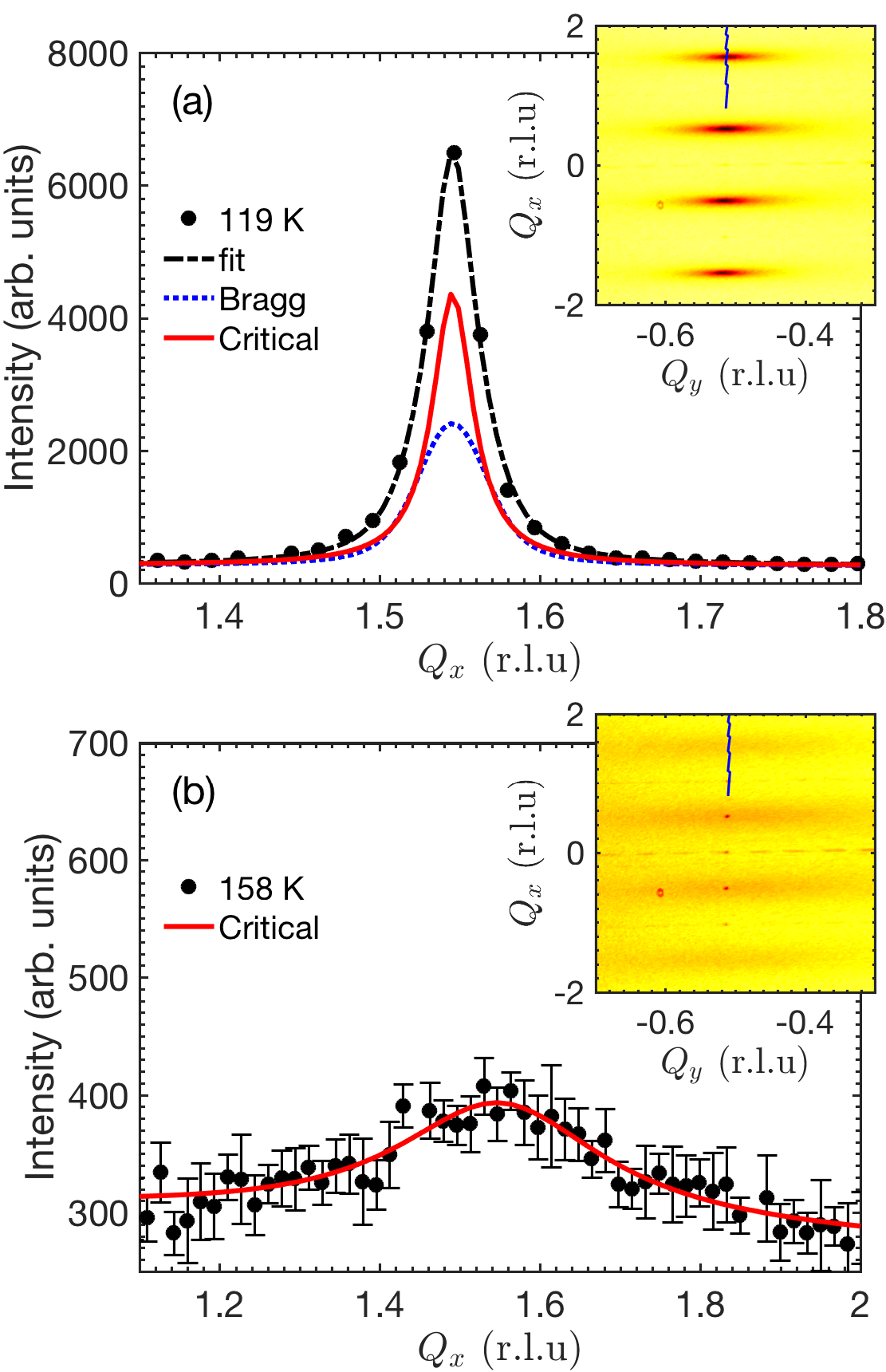}
   	\caption{Critical scattering profiles of an antiferromagnetic Bragg peak at $(Q_x,Q_y)=(\frac{3}{2},-\frac{1}{2})$ taken along the $Q_x$ direction in reciprocal space. The profile from Array 1 (a) at 119 K was fitted with the sum of a Lorentzian and Voigt profile and (b) at 158~K was fitted with a Lorentzian profile only. The insets are the corresponding scattering patterns measured on the CCD camera. The (blue) line represents the portion of reciprocal space used to obtain the profiles shown in the main figure.}
	\label{fig:figure2}
\end{figure}
%=== end figure ==========================%

%%%%%%%%%%%%%%%%%%%%%%%%%%%%%%%%%%%%%%%%%%%%%

\section{\label{sec:rxd}Soft x-ray scattering}

Soft x-ray resonant magnetic scattering is an ideal tool to study phase transitions in thermally active artificial spin systems due to the large resonant magnetic scattering cross section of x-rays and therefore high sensitivity to the very small volume of magnetic material in the sample, and the high brilliance of synchrotron radiation that permits the study of small area nanopatterned arrays. In addition, the intrinsic high resolution of the momentum transfer, \textbf{Q}, offers a precise measurement of the scattering intensity distribution in reciprocal space.\cite{vale2015, granado2004}

Soft x-ray scattering experiments were performed in the RESOXS chamber\cite{staub2008} of the SIM beamline\cite{flechsig2010} at the Swiss Light Source, Paul Scherrer Institute. Experimental scattering patterns were measured using circularly polarized incident x-rays at an energy of 708~eV corresponding to the Fe L$_3$ edge and at 690~eV (below the absorption edge) for the normalization of the patterns taken at the absorption edge. At resonance, the small angle reflection geometry ensures sensitivity to the in-plane magnetic moments of the sample. The angle of incidence of the x-ray beam ($\theta_i=8^{\circ}$) was kept constant throughout the experiment (Fig.~\ref{fig:geometry}). {\color{black}The size of the illuminated region was approximately 100 $\mu m$ by 850 $\mu m$, which is comparable to the size of the arrays themselves of 150 $\mu m$ by 1500 $\mu m$.} The diffraction patterns were acquired using a Princeton Instrument PME charge-coupled device (CCD) camera with 1340$\times$1300~pixels with 20$\times$20~$\mu$m$^2$ pixel size. The Q-resolution of the instrument was determined from the off-resonance structural Bragg peaks and was found to be $5.7\times 10^{-4}$~$nm^{-1}$ perpendicular to the scattering plane (along $Q_x$). While measuring the magnetic scattering, we place a mask in front of the CCD detector to block the high intensity structural Bragg peaks and the specular reflected beam,\cite{sendetskyi2016} so that we were able to follow the weak magnetic critical scattering {\color{black}up to $\sim$90~K above the phase transition}. At higher temperatures, the critical scattering becomes too weak and broad to extract the fit parameters reliably.

In order to follow the evolution of the scattering pattern as a function of temperature, the sample was first slowly cooled from room temperature to 10 K at a rate of about $\approx$~2~K/min and the scattering intensity was then measured across the transition on heating up. For arrays 2 and 3, an additional series of scattering patterns as a function of temperature was collected in order to have sufficient sampling points in the vicinity of the phase transition and, for this, the sample was only cooled down to 170K due to the limited experimental time. Approximately twenty minutes were spent at each temperature, which provided sufficient time to take measurements of all three arrays. At each temperature, {\color{black} approximately ten images were taken. Each image was binned,  combining a clusters of 4 pixels (along the horizontal) by 2 pixels (along the vertical) into a single pixel. Subsequently, the images were normalized to the incident beam intensity measured with a monitor placed in front of the experimental chamber. The resulting normalized images were then stacked in a three dimensional matrix and an averaged image with corresponding error bars was obtained by standard statistical methods.}  

Antiferromagnetic ordering of magnetic moments in the ground state, see Fig.~\ref{fig:figure1}, can be represented by a magnetic lattice rotated by $45^{\circ}$ and periodicity $\sqrt{2}$ times bigger than the structural lattice periodicity. In reciprocal space, pure magnetic peaks will be well separated from the structural ones and have a periodicity of $\frac{2\pi}{a\sqrt{2}}$, where $a$ is the structural lattice periodicity. Examples of scattering patterns in the vicinity of the $(Q_x,-\frac{1}{2})$ family of antiferromagnetic Bragg peaks are shown in the insets of Fig.~\ref{fig:figure2}.

\begin{figure}[t]
\includegraphics[width=0.94\linewidth]{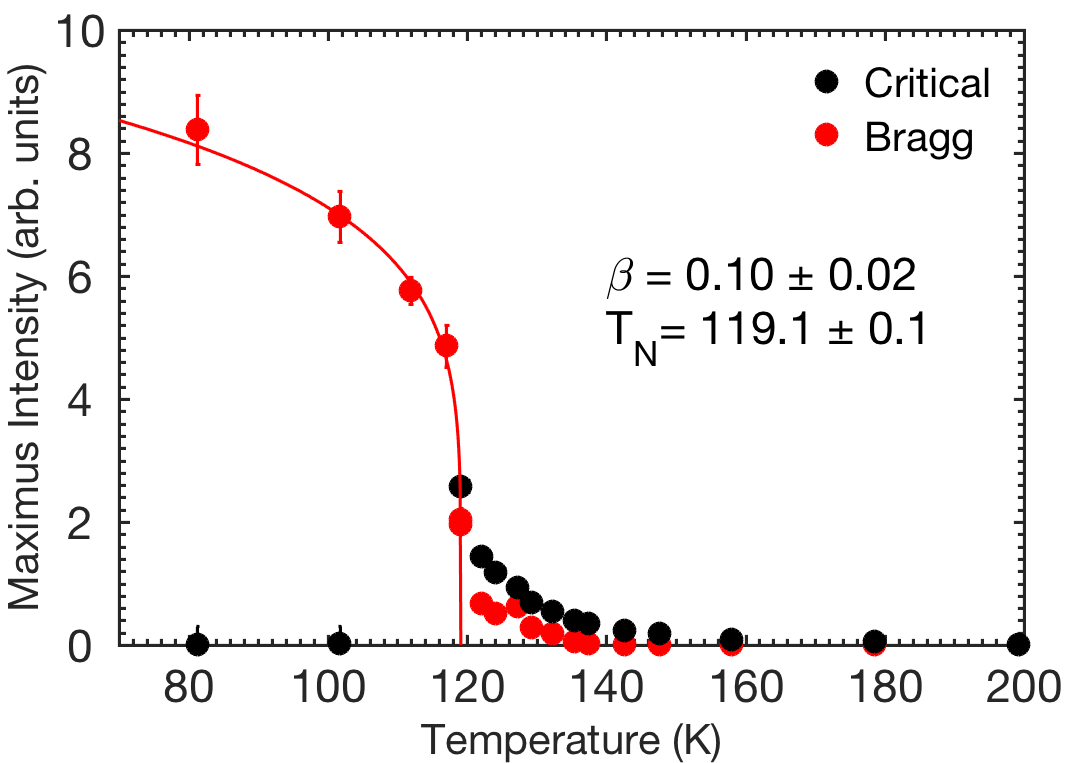} 
       \caption{Intensity of the Bragg peak (red circles) and critical scattering (black circles) extracted by fitting the peak profiles from Array~1. The transition temperature was determined to be $T_{N}=119\pm0.1~K$ using a power law fit (red line). Due to the weakness of the critical scattering just below $T_{N}$, it was not possible to unambiguously separate the two components in the scattered intensities, so that only the Bragg contribution was estimated in the fits.}
       \label{fig:figure3}
      \end{figure}

Peak profiles at the $(\frac{3}{2},-\frac{1}{2})$ position were obtained by relating the pixel position of the two-dimensional images from the CCD camera to reciprocal space values\cite{valerio1}. Due to the geometry of the experiment and the small size of the reciprocal unit cell, it was possible to measure  a large portion of reciprocal space around the diffraction peaks. {\color{black}The scattering profiles were initially fitted with the appropriate standard analytic functions conventionally used in scattering experiments, namely Gaussian, Lorenztian and Voigt profiles. In order to properly describe the observed profiles and to reduce the number of parameters to estimate, we accounted for two contributions to the scattering intensities: (i) the critical scattering represented by a Lorentzian profile (see Ref.~\onlinecite{collins1989}, Chapter 14)  convoluted with a Gaussian profile of fixed width to represent the experimental resolution}, and (ii) the magnetic Bragg scattering represented by a Gaussian profile for Array~2 and Array~3, and a Voigt profile for Array~1. At temperatures far below the Neel transition temperature $T_N$, in the ordered state, the profiles were fitted considering only the Bragg contribution, assuming the critical scattering contribution to be negligible. The width of the Bragg contribution obtained from these fits was kept fixed for all the remaining temperatures. The width of the critical scattering as well as the intensity of the critical and Bragg scattering were kept as free parameters in the fits for each temperature. Representative scattering profiles along the $Q_x$ direction with magnetic Bragg and critical scattering contributions above $T_N$ for Array~1 are shown in Fig.~\ref{fig:figure2}. The intensities obtained for the Bragg and critical scattering contributions for Array~1 are given as a function of temperature in Fig.~\ref{fig:figure3}. The Bragg scattering intensity $I$ shows a strong increase as the transition is approached from above. This corresponds to the emergence of antiferromagnetic order below $T_N$.
%==== figure =============================%
%=== end figure ==========================%

\section{\label{sec:nature}Nature of the phase transition}
%%%%%%%%%%%%%%%%%%%%%%%%%%%%%%%%%%

To establish the nature of the antiferromagnetic transition in artificial square ice, we first look at the basic characteristics of the predicted ground state. The long range ordered ground state of the square ice lattice is shown in Fig.~\ref{fig:figure1}(b) and consists of two sublattices of red (horizontal) and blue (vertical) antiferromagnetically aligned moments. The ground state has a two-fold degeneracy and may be described by an array of vertices defined on a lattice whose points belong to the geometrical center of a plaquette formed by each set of four neighbour Ising magnets. For the ground state, the magnetic state of this plaquette corresponds to two nearest neighbour moments pointing inwards and two pointing outwards, and is referred to as a vertex of Type~I, as shown in Fig.~\ref{fig:figure1}(b). The two-fold degeneracy arises because the ground state can consist of either one of the two possible Type~I vertices, resulting in the tiling of alternating chiral micro-vortices.\cite{morgan2011} At finite temperature, fluctuations will involve the creation of Type~II, III and IV vertices (see Fig.~\ref{fig:figure1}(b) for their definitions). In terms of a net magnetic flux, Types~III and IV (the highest energy vertex configurations) can be viewed as emergent charge excitations.~\cite{moller2009} Both the two antiferromagnetic sublattices and the alternating chirality descriptions of the ground state suggest a long range ordered ground state whose broken symmetry is that of the two dimensional Ising class. Therefore, we expect the presence of a continuous phase transition with this corresponding class at some finite critical temperature.

\begin{figure}[b]
\includegraphics[width=0.92\linewidth]{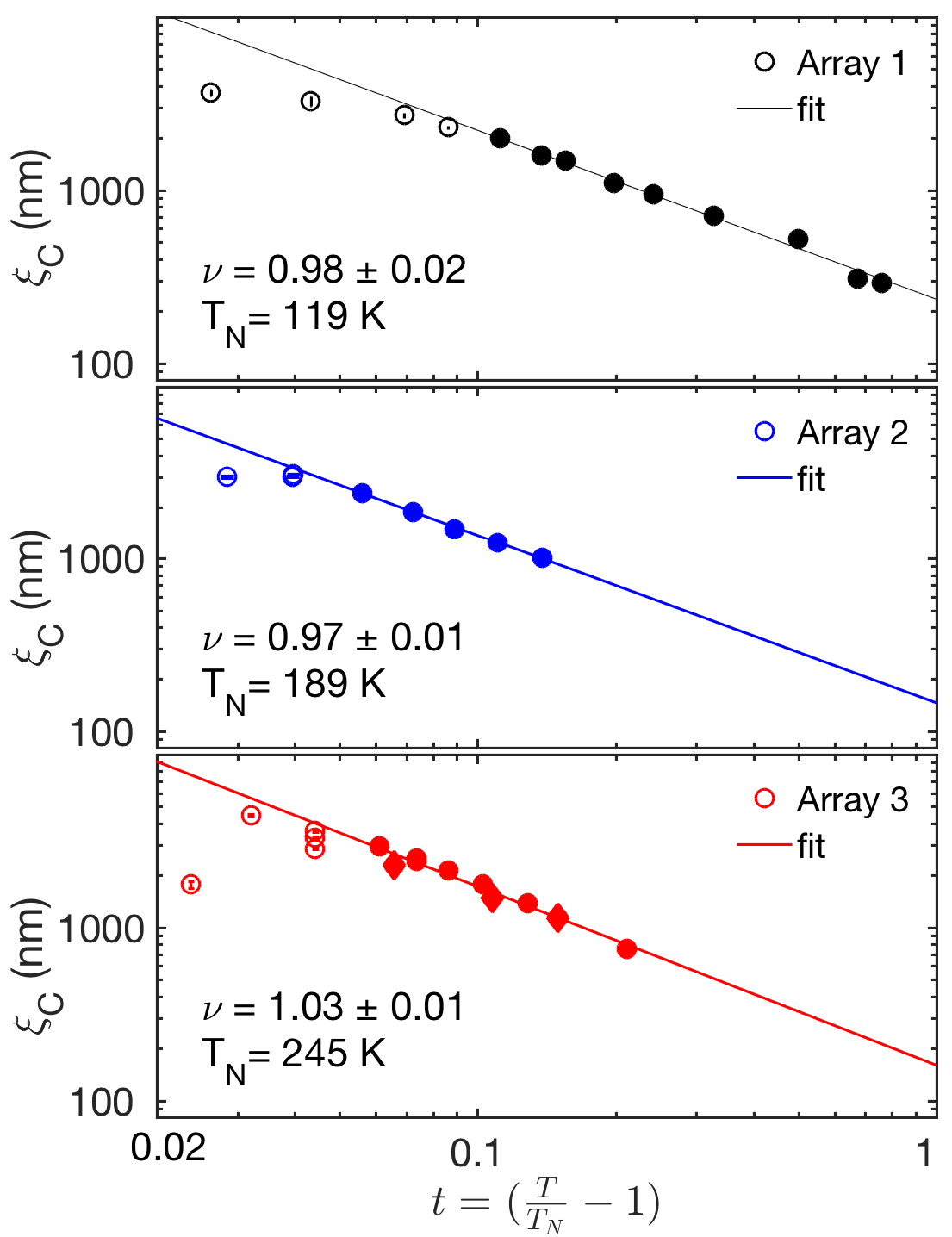} 
       \caption{Temperature dependence of the critical correlation length in the direction perpendicular to the scattering plane for array 1 (black circles), array 2 (blue circles), and array 3 (red circles and diamonds) of artificial square ice. For array 3, data were collected in two different temperature runs, indicated by circles and diamonds, respectively. Solid lines with the corresponding color are fits to a power law giving the critical correlation length exponent $\nu$. The fit was performed using the data points indicated by closed symbols. }
      \label{fig:correlation}
      \end{figure}

In order to experimentally determine the nature of the phase transition, first we extract from the experimental data the transition temperatures $T_N$ and the magnetization exponents $\beta$ for the three arrays using the power law expression $I \propto (-t)^{2\beta}$, where the reduced temperature $t=(T - T_N)/T_N$ and $T<T_N$. The fitted values of $T_{N}$ and $\beta$ for the arrays are given in Table~\ref{tab:table1}. The values for the magnetization exponents are close to the expected value, $\beta=0.125$, for a 2D Ising universality class, although the measurement is not very precise with a relative error in $\beta$ of 30-40\%. Such a large uncertainty prevents a conclusive statement about the universality class of the phase transition. However, a more accurate determination of another critical exponent is possible by considering the critical behaviour of the system above $T_{N}$, where the magnetic Bragg scattering is negligible. Specifically, the critical scattering spatial correlation length $\xi_C$ should follow the scaling relation $\xi_C \propto (t)^{-\nu}$ near $T_{N}$, with $\nu$ being the correlation length exponent, which is expected to be equal to 1 for the 2D Ising universality class and $T>T_N$. $\xi_C$ is readily obtained, at each temperature, from the full-width-at-half-maximum (FWHM) of the critical scattering peak by using the fitting procedure outlined in the Section~\ref{sec:rxd}. Indeed, $\xi_C$ is proportional to the inverse of the FWHM of the peak. The dependence of $\xi_C$ versus the reduced temperature $t$ is given for all of the arrays in Fig.~\ref{fig:correlation}. In Fig.~\ref{fig:correlation} only temperatures above $T_{N}$ are shown ($t=0$ at $T_N$), since it is only in this regime that the scaling relation could be reasonably well fitted. Indeed, above $T_{N}$, an accurate value of the critical exponent $\nu$ could be found with a relative error of about 2\%. 

While the estimated exponents for {\color{black}all arrays are very close to the theoretical value for the 2D Ising universality class ($\nu=1$),}
%
%
%==== table =============================%
\begin{table*}
\centering
\color{black}\begin{tabular}{|c|c|c|c|c|c|c|c|c|}
\hline
 & $h$ & $T_N$ & $T_B$ & $T_N$-$T_B$ & $\xi_S(LRO)$ $\pm2\%$& $\frac{2\pi}{\xi_S(LRO)}$ $\pm2\%$ & $\beta$ & $\nu$ \\
\hline
Array 1 & 4.1~nm & 119.1 $\pm0.1$ K & 68 $\pm2 $ K & 51 $\pm2 $ K & 4020 nm ($\approx$ 24 u.c.) & $1.56 \times 10^{-3}$ $nm^{-1}$ & 0.10 $\pm$ 0.02 & 0.98 $\pm0.02$ \\
\hline
Array 2 & 4.3~nm & 189 $\pm2 $ K & 85 $\pm 3 $ K & 104 $\pm 5 $ K  & 7380 nm ($\approx$ 43 u.c.) & $0.85 \times 10^{-3} nm^{-1}$ & 0.11 $\pm$ 0.04 & 0.97	 $\pm 0.01$  \\
\hline
Array 3 & 4.5~nm & 245 $\pm 2$ K & 100 $\pm 3 $ K & 145 $\pm 5 $ K & 8700 nm ($\approx$ 51 u.c.) & $0.72 \times 10^{-3} nm^{-1}$ & 0.13 $\pm$ 0.05 & 1.03	$\pm 0.01 $  \\
\hline
\end{tabular}
\caption{\label{tab:table1}Summary of Permalloy thicknesses $h$, {\color{black}transition temperatures $T_N$, blocking temperatures $T_B$, their difference $T_N-T_B$, static correlation lengths in the long range ordered phase $\xi_S(LRO)$ and $\frac{2\pi}{\xi_S(LRO)}$, and critical exponents $\beta$ and $\nu$ for each Array.} ``u.c.'' stands for ``unit cells''}
\end{table*}
%====end table =============================%
%
%
 we observe a saturation of the critical correlation length $\xi_C$ as the temperature approaches the transition temperature $T_{N}$ ($t=0$). The origin of this plateau is related to the closeness of $T_{N}$ to the blocking temperature of artificial spin ice systems and will be discussed in more detail in Sec.~\ref{sec:effects}. 

Finally, we note that, from the experimental data, it was also possible to extract the low-temperature static correlation lengths, $\xi_S$ (see  Table~\ref{tab:table1}). These were determined from the magnetic Bragg peak FWHM at the lowest temperature, where the critical scattering contribution is expected to be negligible compared to the one originating from the long range magnetic Bragg order. Interestingly, we find that the widths of magnetic peaks in the ordered state ($\approx7.2\times 10^{-4}$~$nm^{-1}$ given by $\frac{2\pi}{\xi_S(LRO)}$ for Array~3) for all the three arrays are bigger than the width of the line shape of the resolution function ($\approx5.7\times 10^{-4}$~$nm^{-1}$, see Sec. \ref{sec:rxd}). We interpret this observation as a fingerprint of the presence of magnetic domains. Finite size clusters of the ordered phase are formed at the second order phase transition and these clusters form magnetic domains below $T_N$. In addition, the system freezes before it can form a single magnetic domain that spans the full array size.

To summarize, soft x-ray magnetic scattering of our artificial square ice system gives critical exponents $\beta\approx0.10-0.13$ and $\nu\approx0.97-1.03$ (see Tab.~\ref{tab:table1}) for the continuous phase transition to low temperature long range order, which are compatible with the Ising universality class exponents of $\beta=0.125$,  $\nu=1$.
 %==== figure =============================%
%=== end figure ==========================%
It should be pointed out that the exponents we have found are consistent with past theoretical work, which has considered the square ice system within the framework of a 16-vertex model.\cite{levis2013,foini2013,levis2013defects}. This simplified model, which truncates the dipolar interaction to include only the nearest neighbour interaction, also admits a long-range ordered antiferromagnetic phase with an exponent of $\eta=1/4$, and the model is found to be in good agreement with magnetic force microscopy measurements on artificial square ice.\cite{levis2013} Further theoretical modelling on this system has also taken into account the quasi long range nature of the dipolar interaction in two dimensions.\cite{silva2012} Here, a logarithmic scaling of the specific heat with respect to system size was predicted, suggesting a two-dimensional Ising exponent of $\alpha=0$. In addition, it was suggested that there is a diverging length scale for type III vertices close to the transition temperature. These two independent works, together with our scattering data, corroborate the idea that the phase transition to the long-range ordered state belongs to the two-dimensional Ising universality or a similar class. 

To test the assumption of a two dimensional Ising universality class, we perform Monte Carlo simulations on the dipole-interacting square ice. In these simulations, each nanomagnet is represented as a point dipole, which has been found to describe well the behaviour of thermally active artificial kagome ice\cite{farhan2013} and square ice\cite{farhanprl2013} systems. The details of the simulation are given in the Appendix. We performed a standard technique of finite size scaling collapse\cite{cardy2012} using the simulation results from several square arrays with increasing dimensions. The sizes of the systems is specified by $n=L/a=8$, 10, 16, 20, and 40, where $L$ corresponds to the side length of the simulated array. {\color{black}It is worth mentioning here that the size of the experimental arrays of nanomagnets is orders of magnitude larger than the size that can currently be simulated due to computational constraints. Therefore, for transitions in more complex systems, such as highly frustrated systems or systems with either discrete or continuous degrees of freedom, where the size of the system used in the simulations is of utmost importance, experiments might become a deciding factor for elucidating the nature of the observed phase transition.} 

\begin{figure}[h!]
	\includegraphics[clip,width=0.99\linewidth]{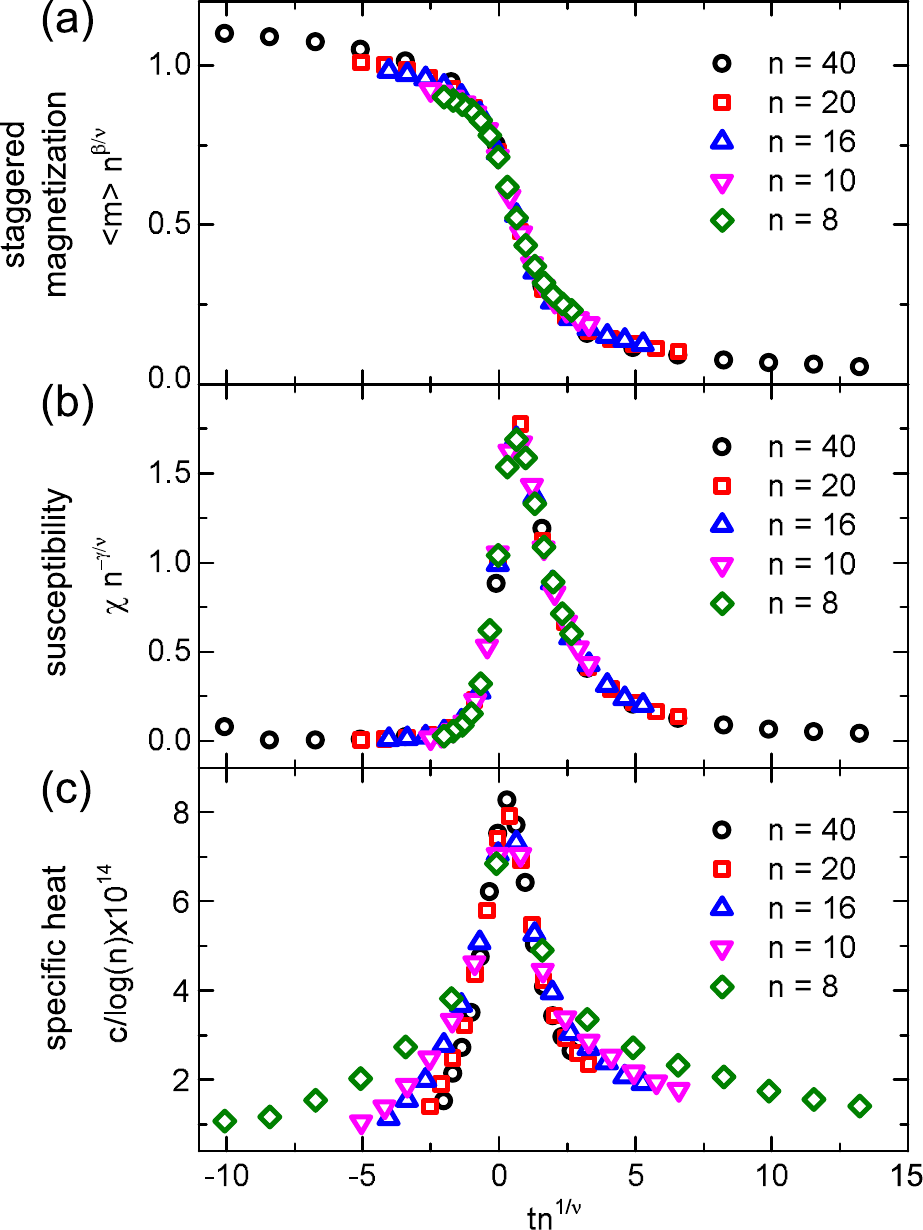}
	\caption{(a) Staggered magnetization per site, (b) magnetic susceptibility and (c) specific heat scaling collapse assuming critical exponents of two dimensional Ising system.}
	\label{fig:figure7}
\end{figure}

The magnetization, susceptibility and specific heat are calculated for each system size and the finite size scaling plots are given in Fig.~\ref{fig:figure7}.\cite{landau2014} Such scaling collapses are obtained by scaling the thermodynamic data with respect to powers of the system length scale (here denoted by $n$). Therefore, the reduced temperature $t$ is multiplied by $n^{1/\nu}$, the staggered magnetization is divided by $n^{\beta/\nu}$, the susceptibility is divided by $n^{\gamma/\nu}$ and the specific heat is divided by $\log n$. In the plots, the critical exponents of the two dimensional Ising universality class are assumed: $\nu=1$, $\beta=0.125$, $\gamma=7/4$ (susceptibility exponent), $\eta=1/4$ (correlation function exponent) and $\alpha=0$ (specific heat exponent). The scaling collapse shown in Fig.~\ref{fig:figure7} is very good, with all curves lying on top of each other, indicating that the behavior at the transition in artificial square ice is similar to that of the two dimensional Ising universality class.

Thus, both experiment and simulation demonstrate a transition to low temperature long range order whose behaviour is very similar to a 2D Ising universality class transition. 

 %==== figure =============================%
 %==== end figure =============================%

%%%%%%%%%%%%%%%%%%%%%%%%%%

\section{\label{sec:effects}Tuning of the critical and blocking temperatures}

In this section, we discuss the possibility to tune the critical behaviour of artificial square ice by changing the parameters of the nanomagnets, in this case their thickness. We start with a unique feature of thermally active artificial spin systems, namely that a single-domain elongated nanomagnet will have a temperature-dependent timescale of the magnetic moment fluctuations between two degenerate states, analogous to the temperature-dependent timescale of the fluctuations of atomic spins in the Langevin paramagnetism theory.\cite{bean1959}  Therefore there exists a temperature $T_B$, commonly referred to as the blocking temperature, below which the moments of the nanomagnets will stop fluctuating and become static at the timescale of the experiment. Thus at temperatures below $T_B$, the system cannot achieve equilibrium within the timescale of the experiment.

Assuming the reorientation timescale is given by an Arrhenius form, $\tau_{\mathrm{reor}}=\tau _{0}\exp \left[KV/(k_{B}T)\right]$, where $k_B$ is the Boltzman constant, $V$ is the volume of the nanomagnet, $K$ is the anisotropy constant, and $\tau_{0}$ is the moment reorientation attempt rate, the blocking temperature is the temperature at which $\tau_{\mathrm{reor}}$ equals the time-scale of the experiment ($\tau_{\mathrm{exp}}$). That is, $T_B$ is defined as 
\begin{equation}
T_{B}={\frac  {KV}{k_{B}\ln \left({\frac{\tau_{\mathrm{exp}}}{\tau_{0}}}\right)}}={\frac  {K w l h}{k_{B}\ln \left({\frac{\tau_{\mathrm{exp}}}{\tau_{0}}}\right)}}. \label{bt}
\end{equation}
In the above, $w$, $l$ and $h$ are the width, length and thickness of the nanomagnet as defined in Sec.~\ref{sec:sample}.  Assuming $\tau_{0}\simeq1.67\times10^{-11}$ minutes~\cite{Liashko2017} and $\tau_{\mathrm{exp}}\simeq20$ minutes (see Sec.~\ref{sec:rxd}), we obtain $\ln(\tau_{\mathrm{exp}}/\tau_{0})\simeq28$. The dependence of the $T_B$ on the nanomagnet parameters in \eqref{bt} should be compared with the $T_N$ dependence, given by
\begin{equation}
T_N \propto m^2 = V^2 M_s^2 = w^2 l^2 h^2 M_s^2,
\label{tn}
\end{equation}
where $m$ is the moment of the nanomagnet. Thus $T_B \propto h$ and $T_N \propto h^{2}$ and, by increasing (reducing) the thickness of the nanomagnets, one can achieve a larger (smaller) separation between $T_B$ and $T_N$, assuming $T_B$ is lower than $T_N$. 

\begin{figure}
\includegraphics[width=0.9\linewidth]{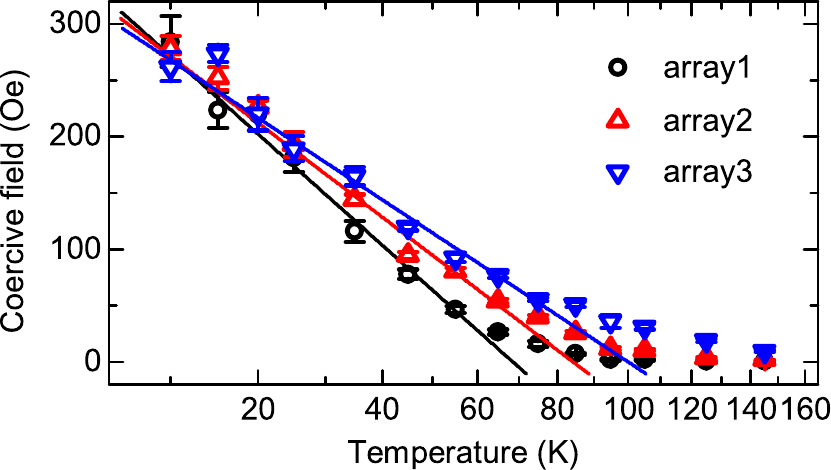} 
       \caption{Temperature dependence of the coercive field $H_c$ for Array~1 (black circles), Array~2 (red up-pointing triangles) and Array~3 (blue down-pointing triangles) obtained from magneto-optical Kerr effect (MOKE) measurements, displayed using a square root scale. Solid lines correspond to the fits performed using  Eq.~\eqref{blocking} to obtain the single-particle blocking temperature.}
       \label{fig:figure10}
      \end{figure}

The blocking temperature $T_B$ of each array was obtained via temperature-dependent magnetization curves measured with magneto-optical Kerr effect (MOKE) magnetometry with a magnetic field applied along the [1~0] lattice direction, see Fig.~\ref{fig:figure1}. The blocking temperature was then estimated from the temperature-dependent coercivity of each array, which is described by the following equation\cite{bean1959}
\begin{equation}
H_c=H_0 \left[1-\left(\frac{T}{T_B}\right)^\frac{1}{2}\right].
\label{blocking}
\end{equation}
The MOKE hysteresis loops were measured at different temperatures going from 10~K to 160~K in steps of 10~K. The time spent at each temperature was 40-45~minutes, which is comparable to the time used for each x-ray measurement. The measured temperature dependent coercive fields were fitted using Eq. \eqref{blocking} and are plotted in Fig.~\ref{fig:figure10}. The resulting blocking temperatures $T_B$ for the arrays investigated in our experiment are summarized in Table~\ref{tab:table1}. We find that a thickness increase from one array to another of 0.2~nm changes $T_B$ by $\sim15$~K. For Array~1, $T_N$ - $T_B$ $\approx$ 51~K, whereas $T_N$ - $T_B$ $\approx$ 104~K and 145~K for Array~2 and Array~3, respectively. 

%
%
 %==== figure =============================%
%=== end figure ==========================%
%
%
If $T_B<T_N$, our square-ice system remains thermally active at $T_N$. However, as the temperature approaches $T_N$ from above and the correlation length diverges, as predicted for a continuous phase transition, larger domains of order emerge and the timescale at which their structure non-negligibly changes increases. This is referred to as critical slowing down and has the effect that any system will fall out of equilibrium in the temperature interval between $T_{f-}$ and $T_{f+}$, the so-called freeze-out region, where $T_{f-}$ and $T_{f+}$ are the freeze-out temperatures below and above $T_N$ respectively. It should be mentioned that critical slowing down is a collective effect, which is distinct from the single-particle blocking associated with the blocking temperature $T_B$, although they both drive the system out-of-equilibrium. For the case of a microscopic system where $T_B$ is orders of magnitude smaller than $T_N$, critical slowing down can be detected only when the temperature quench from above is fast enough. For our artificial square ice arrays, the blocking and transition temperatures are of the same order of magnitude. As a consequence, very strong critical slowing down is expected and, on cooling down, the freeze-out temperature $T_{f+}$ will be further away from the transition temperature. Indeed, we see in Fig.~\ref{fig:correlation} that the critical correlation length $\xi_C$ {\color{black}diverges from the expected critical behaviour}  for all arrays before reaching $T_N$ (when $t=0$). The increase of $T_{f+}$ from Array~3 to Array~1 can be seen from the critical correlation length $\xi_C$ plots shown in Fig.~\ref{fig:correlation} where for Array~1, the deviation from power law-like behaviour is seen at higher reduced temperature compared to the other two arrays.  {\color{black}For Array~1, for which the blocking and transition temperatures differ by only $\approx51$~K, the deviation is more significant, due to the higher freeze-out temperature $T_{f+}$}. Similarly, the static correlation length $\xi_S$, which is proportional to the magnetic domain size in the ordered phase, is higher for Array~2 and Array~3 compared to Array~1 (see Table~\ref{tab:table1}), {\color{black}but is, in any case, well above the instrumental resolution}.  Such observations confirm the expectation of a strong out-of-equilibrium dynamics occurring in our system in the vicinity of the phase transition.

In the above, we have discussed how the single-particle blocking modifies the freeze-out temperature $T_{f+}$ above the transition temperature $T_N$. If we now look at what happens below the phase transition, we notice that there is an increase in the domain size in the antiferromagnetic phase, estimated from $\xi_S$, as the value of $T_N-T_B$ increases, see Table~\ref{tab:table1}. This behaviour indicates that the single-particle blocking facilitates the freezing of the domain structure, or more generally speaking the defect density, below the phase transition and effectively increases the value of $T_N - T_{f-}$ compared to $T_{f+} - T_N$. 

The formation of defects and their density at a continuous phase transition are generally described by the Kibble-Zurek mechanism.~\cite{del2014,zurek1996,griffin2012,beugnon2017} {\color{black} While numerical calculations suggest that pyrochlore spin ice materials are well-suited for experimental investigation of Kibble-Zurek scaling due to their slow spin dynamics,\cite{hamp2015} there is also increasing interest in out-of-equilibrium dynamics in theoretical models of artificial spin ice.\cite{levis2012}} In experimental artificial spin systems, the vicinity of $T_B$ to $T_N$ has the additional effect of slowing down the critical fluctuations below $T_N$ and can help to preserve the domain structure formed at the phase transition. While at first sight the single-particle blocking might look like an unwanted complication in studying the critical properties of the system, using artificial spin systems could be advantageous for testing the Kibble-Zurek mechanism and out-of-equilibrium dynamics of phase transitions. {\color{black} In particular, a common method of measuring Kibble-Zurek scaling is to perform temperature quenches of the sample across the phase transition at different rates, which is often experimentally challenging. For artificial spin systems, one could instead manufacture a sample with arrays of nanomagnets with varying single-particle blocking temperature, and therefore with different freeze-out temperatures, and measure the scattering patterns at the same time at a single, experimentally convenient, quench rate. In this way, the formation of non-trivial topological defects could be studied under controlled conditions.}

{\color{black}  To achieve this, the most elegant approach is to create several nanomagnet arrays with different thicknesses from a Permalloy thin film wedge on a single substrate, as carried out in the present work. In this way, both $T_B$ and $T_N$ can be modified in a desired manner by changing only one parameter. One could also further reduce $T_B$ by reducing the shape anisotropy (by changing the lateral geometry of the nanomagnets) and increase $T_N$ by reducing the distance between the nanomagnets or using a material with higher saturation magnetization $M_S$.}

\section{\label{sec:conclusions}Conclusions}

In conclusion, we have obtained a quantitative signature of a continuous second order phase transition in artificial spin ice, extending the concepts of criticality and universality class to mesoscopic spin systems. Critical exponents determined from soft x-ray resonant magnetic scattering experiments and the finite size scaling collapse obtained from Monte Carlo simulations indicate that the transition in artificial square ice belongs to the two-dimensional Ising universality class. {\color{black}With the emergence of magnetic criticality in artificial spin ice, we have demonstrated that complex many-body phenomena can be investigated in nanomagnetic systems using modern synchrotron x-ray scattering methods, which provide a powerful alternative to imaging methods.} By taking advantage of the flexibility to vary the nanomagnet parameters of artificial spin ice, we have also demonstrated the role of out-of-equilibrium dynamics that originate from the proximity of the blocking temperature to the transition temperature. These effects are exemplified by a {\color{black}significant deviation from the expected universality class power law behaviour in the vicinity of T$_N$}.  

Our results have three major implications for the elucidation of the nature of phase transitions. {\color{black}First, the ability to measure critical exponents in artificial spin systems means that it would be possible to experimentally investigate the two-dimensional antiferromagnetic Ising model under an applied magnetic field.\cite{lourencco2016} Indeed, artificial square ice should admit a number of antiferromagnetic phases at finite field whose character depends strongly on the value of both nearest and next nearest neighbor coupling strengths.\cite{binder1980,lourencco2016}} Second, in order to ascertain the critical behaviour at phase transitions, one should carefully design artificial spin systems to ensure that $T_B$ is far enough away from $T_N$. {\color{black}As we have shown, one of the ways to ensure that $T_B$ is well below $T_N$ is to make the nanomagnets thicker, while ensuring that the equilibration time is within the timescale of the experimental measurement. This approach is also relevant for observing the phase transitions in artificial kagome spin ice with scattering, imaging or spectroscopy techniques.\cite{sendetskyi2016, anghinolfi2015, farhan2013, perrin2016}} Thirdly, using artificial spin systems, one can explore experimentally out-of-equilibrium dynamics close to the phase transition governed by the Kibble-Zurek mechanism.\cite{zurek1996} This would allow one to predict domain size and charge defect density as the temperature approaches the critical temperature.

While there are several one- and two-dimensional mesoscopic systems that have experimentally accessible phase transitions, such as trapped atomic gases,\cite{hadzibabic2006} liquid crystals,\cite{chuang1991} ultrathin magnetic films\cite{rose2007} and interacting Josephson junctions,\cite{resnick1981} artificial spin systems particularly lend themselves to a more detailed investigation of out-of-equilibrium dynamics governed by the Kibble-Zurek mechanism in condensed matter physics, as it is possible to tune the intrinsic magnetization dynamics by changing the blocking temperature. Using the possibility to tune the blocking temperature, one could not only determine the dependence of the population of magnetic charge defects and domain size on the cooling rate, but also the dependence on $T_N-T_B$.

{\color{black}As an outlook, we believe that modern scattering methods can also be successfully used to answer fundamental questions about the nature of phase transitions and dipolar interactions in artificial spin systems with continuous degrees of freedom, such as XY systems,\cite{ewerlin2013,vedmedenko2005} and other highly frustrated systems,\cite{smerald2016,smerald2017} which have proven to be hard to tackle both numerically and experimentally. Furthermore, our approach will complement the local real-space probes such as photoemission electron microscopy and magnetic force microscopy, and will help to understand how charge defects, vertex chirality and modifications to the geometry and magnetic interactions \cite{gliga2017,gliga2015,ostman2017,perrin2016} leads to different critical properties and universality classes of artificial spin systems, for example, by applying the methods described in our manuscript to samples with broken symmetry along the out-of-plane direction.\cite{perrin2016} Finally, this work will promote extensive future research on determining the nature of exotic phases, such as emergent Coulomb phases, \cite{henley2010,canals2016,brooks2014}that can exist in highly frustrated pyrochlore and artificial spin ices.}

$Note$:
the data that support this study are available via the Zenodo repository (10.5281/zenodo.3236819)

\begin{acknowledgments}
We are grateful to Vitaliy~Guzenko, Anja~Weber, Thomas~Neiger, Michael~Horisberger and Eugen~Deckardt for their support in the sample preparation, Pascal~Schifferle for technical support at the X11MA beamline, Joachim Kohlbrecher for the support of the project and Quintin Meier for helpful discussions. Part of this work was carried out at the X11MA beamline of the Swiss Light Source, Paul Scherrer Institute, Villigen, Switzerland. The research leading to these results has received funding from the Swiss National Science Foundation and the European Community's Seventh Framework Programme (FP7/2007-2013) -- Grant No.~290605 (COFUND PSI-FELLOW).
\end{acknowledgments}

\appendix*

\section{\label{appendix}Monte Carlo simulations} 

All simulations are performed under periodic boundary conditions using lattices with a square geometry defined by the number of Bravais lattice vectors $n$ along the $\mathbf{\hat{x}}$ and $\mathbf{\hat{y}}$ directions. The number of Ising spins is therefore equal to $N=2n^{2}$.  The square-ice lattice is constructed using a conventional cell defined by the orthogonal vectors $\mathbf{a}_{1}=a\mathbf{\hat{x}}$ and $\mathbf{a}_{2}=a\mathbf{\hat{y}}$. Here $a$ is the lattice constant defined by the distance between neighboring parallel Ising magnets. Each conventional cell contains two Ising spins whose locations within the cell are given by the vectors $\mathbf{a}_{1}/2$ and $\mathbf{a}_{2}/2$, with respective Ising axis directions $\mathbf{\hat{x}}$ and $\mathbf{\hat{y}}$. Each Ising magnet is modeled as a point dipole with moment direction $\mathbf{\hat{m}}_{i}$ and magnitude $m=VM_{\mathrm{s}}$. Here $V$ is the volume of the Ising magnet and $M_{\mathrm{s}}$ is the saturation magnetization.  The present simulations use $a=168$ nm, a volume $V=l\times w\times h=118\times47\times3$ nm$^{3}$ defined by the dimensions of the nanomagnet, and $M_{\mathrm{s}}=200\time10^{-3}$ A/m. A lower value of saturation magnetization of the nanomagnets compared to the bulk Permalloy saturation magnetization is used following Ref. \onlinecite{farhan2013}.

The interaction energy between two point dipoles is given by the dipolar interaction
\begin{equation}
V_{ij}= -\frac{\mu_{0}m^{2}}{4\pi r_{ij}^3} [3(\mathbf{\hat{m}}_i\cdot\hat{\mathbf{r}}_{ij})(\mathbf{\hat{m}}_j \cdot \hat{\mathbf{r}}_{ij}) -\mathbf{\hat{m}}_i\cdot\mathbf{\hat{m}}_j].
\label{eqn1}
\end{equation}
To ensure convergence with respect to the range of the dipolar interaction, image cell summations are used to include interaction distances up to 40$\times a$ for all considered system sizes. A broad range of temperatures were investigated with particular emphasis on $T>T_N$. For each temperature, one million Monte Carlo sweeps were performed to obtain well-converged thermodynamic and order parameter quantities. 

\begin{figure}[b]
	\includegraphics[clip,width=0.95\linewidth]{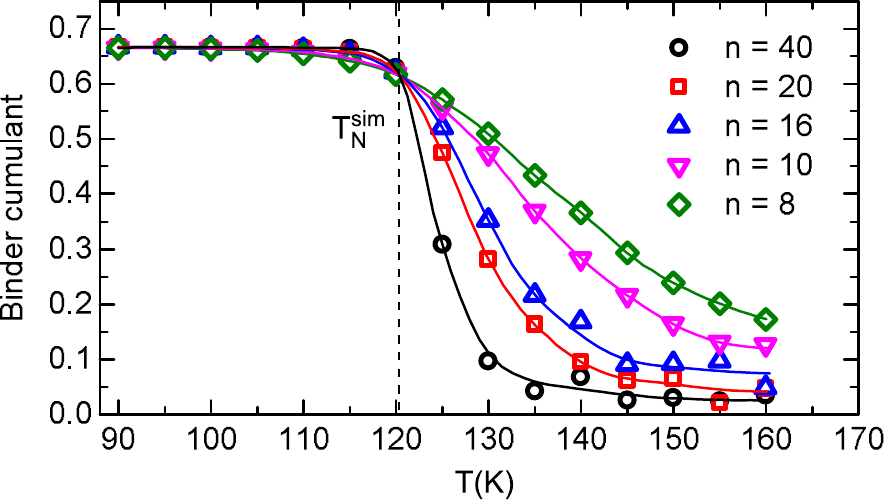}
	\caption{Binder cumulant as a function of temperature for a range of system sizes. Solid lines are guides to the eye.}
	\label{fig:figure6}
\end{figure}

To investigate the temperature dependence of the magnetization dynamics, the magnetic structure factor $\mathbf{M}_{\mathbf{Q}}$ is used. $\mathbf{M}_{\mathbf{Q}}$ is defined as the Fourier transform of the magnetic configurations 
\begin{equation}
\mathbf{M}_{\mathbf{Q}}=\frac{1}{\sqrt{N}}\sum_{j=1,N}
\mathbf{\hat{m}}_{j}\exp\left(i\mathbf{r}_{j}\cdot\mathbf{Q}\right). \label{EqnA2}
\end{equation} 
In the above, $\mathbf{Q}$ can take on discrete values $(\frac{2\pi}{na}n_{x},\frac{2\pi}{na}n_{y})$ where $n_{x}$ and $n_{y}$ are integers ranging between $-n/2$ and $n/2$. The magnetic structure factor calculated at the antiferromagnetic peak, $\mathbf{Q}_{\mathrm{AFM}}=(\pm\pi/a,\pm\pi/a)$, is used as the order parameter.
In particular 
\begin{equation}
M^{2}=\frac{1}{2}\sum_{\mathbf{Q}_{\mathrm{AFM}}}S(\mathbf{Q}_{\mathrm{AFM}})=
\frac{1}{2}\sum_{\mathbf{Q}_{\mathrm{AFM}}}\left|\mathbf{M}_{\mathbf{Q}}\right|^{2}, \label{EqnA3}
\end{equation}
which gives the staggered magnetization per site ($\langle m \rangle=\langle \sqrt{M^{2}} \rangle/N$), the susceptibility ($\chi = (\langle M^{2}\rangle-\langle M\rangle^{2})/T$) and the Binder cumulant\cite{binder1981} ($1-\langle M^{4}\rangle/\langle M^{2}\rangle^{2}/3$). In addition to these thermodynamic quantities, the specific heat (defined as c=$\langle\delta U^{2}\rangle/(T^{2}N)$) is also calculated.  These quantities are determined via single-site ensemble Monte Carlo simulations as a function of both temperature and system size using the experimentally determined values of $m$ and $a$. For the present work, the system sizes are $n=L/a=8$, 10, 16, 20, and 40, where $L$ corresponds to the side length of the simulated array. The Binder cumulant as a function of temperature for the considered system size is given in Fig.~\ref{fig:figure6}. In this figure, the temperature corresponding to the point at which all the curves cross provides an estimate of the critical temperature for an infinite system\cite{binder1981}, giving a value of $T_{\mathrm{N}}^{\mathrm{sim}}\simeq120$ K.

 %==== figure =============================%
 %==== end figure =============================%
 
\begin{figure}[h!]
	\includegraphics[clip,width=0.9\linewidth]{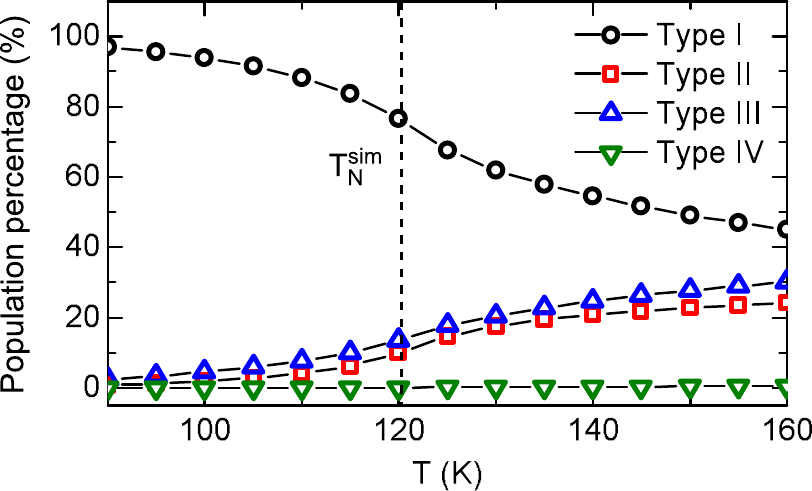}
	\caption{Population percentage of vertex types as a function of temperature for a system size of $n=40$.}
	\label{fig:percentage}
\end{figure}

The low temperature equilibrium populations of each of the four vertex types for the $n=40$ system size are shown in Fig.~\ref{fig:percentage}. The shape of the curves for all vertex types is similar to that arising from the 16-vertex model.\cite{levis2013} At low enough temperatures, the system enters a phase dominated entirely by vertices of Type I. Inspection of the actual spin configurations as the temperature nears zero reveals the ground state of the square-ice system shown in Fig.~\ref{fig:figure1}. In our simulations, the populations of Type~II and Type~III vertices rapidly drop on cooling down to below the critical transition temperature. The population of charge defects associated with Type IV vertices is negligible throughout this regime of temperatures. However, the population of charge defects associated with Type III vertices is not. This latter aspect has been noted in the work of Ref.~\onlinecite{silva2012}, who observe a diverging Type~III vertex (charge) correlation length as $T\rightarrow T_{\mathrm{N}}^{\mathrm{sim}}$. By viewing the ground state as two antiferromagnetic sub-lattices, some insight can be gained into this observation. In fact, the point dipoles in each antiferromagnetic sub-lattice largely follow the ones of the other sub-lattice at temperatures below $T_{\mathrm{N}}^{\mathrm{sim}}$. Spatial regions where this does not occur correspond to domain wall structures which contain a dilute population of Type~III vertices (see Ref.~\onlinecite{budrikis2012, farhanprl2013}). From this perspective, the observed increase in the spatial correlation length between Type~III vertices close to $T_{\mathrm{N}}^{\mathrm{sim}}$, seen by Silva~{\em et al.},\cite{silva2012} would correspond to the growing correlation length associated with critical fluctuations in our simulations, as $T\rightarrow T_{\mathrm{N}}^{\mathrm{sim}}$.
 
   %==== figure =============================%
 %==== end figure =============================%

%\bibliography{os_square}

\begin{thebibliography}{68}
\bibitem [{val()}]{valerio1}%
  \BibitemOpen
  \href@noop {} {}\bibinfo {note} {Note that the deviation of the Bragg peak
  position in Fig. 3 from the expected half integer value is due to the
  impossibility to check the position of the direct beam on the CCD camera,
  which introduces an offset in the peak positions most distance from the
  scatetring plane.}\BibitemShut {Stop}%
\end{thebibliography}

\end{document}